\documentclass[conference]{IEEEtran}
\IEEEoverridecommandlockouts
\pdfoutput=1

\usepackage{geometry}

\usepackage{amsmath,amssymb,amsfonts}
\usepackage{algorithmic}
\usepackage{graphicx}
\usepackage{textcomp}
\usepackage{xcolor}
\usepackage[utf8]{inputenc}
\usepackage[numbers,sort]{natbib}
\usepackage{booktabs}
\usepackage{multirow}
\usepackage{natbib} 

\def\BibTeX{{\rm B\kern-.05em{\sc i\kern-.025em b}\kern-.08em
    T\kern-.1667em\lower.7ex\hbox{E}\kern-.125emX}}

\title{Fabric Sensing of Intrinsic Hand Muscle Activity}

\author{Katelyn Lee\textsuperscript{1},\ Runsheng Wang\textsuperscript{1},\ Ava Chen\textsuperscript{1},\ Lauren Winterbottom\textsuperscript{2},\ Ho Man Colman Leung\textsuperscript{3}, \\ Lisa Maria DiSalvo\textsuperscript{3},\ Iris Xu\textsuperscript{3},\ Jingxi Xu\textsuperscript{3},\ Dawn M. Nilsen\textsuperscript{2,4},\ Joel Stein\textsuperscript{2,4},\ Xia Zhou\textsuperscript{3,4} and Matei Ciocarlie\textsuperscript{1,4} %
\thanks{This work was supported in part by the National Institutes of Health (R01NS115652, F31HD111301), National Science Foundation (IIS-2202553), and a Columbia RISE award.}%
\thanks{$^{1}$K. Lee, R. Wang, A. Chen, and M. Ciocarlie are with the Department of Mechanical Engineering, Columbia University, New York, NY 10027, USA. {\texttt{\footnotesize \{katelyn.lee, matei.ciocarlie\}@columbia.edu}}}%
\thanks{$^{2}$L. Winterbottom, D. M. Nilsen, and J. Stein are with the Department of Rehabilitation and Regenerative Medicine, Columbia University, New York, NY 10032, USA. }%
\thanks{$^{3}$C. Leung, L.M. DiSalvo, I. Xu, J. Xu, X. Zhou, are with the Department of Computer Science, Columbia University, New York, NY 10027, USA. }%
\thanks{$^{4}$Co-Principal Investigators}
}

\begin{document}
\maketitle
\begin{abstract}
    Wearable robotics have the capacity to assist stroke survivors in assisting and rehabilitating hand function. Many devices that use surface electromyographic (sEMG) for control rely on extrinsic muscle signals, since sEMG sensors are relatively easy to place on the forearm without interfering with hand activity. In this work, we target the intrinsic muscles of the thumb, which are superficial to the skin and thus potentially more accessible via sEMG sensing. However, traditional, rigid electrodes can not be placed on the hand without adding bulk and affecting hand functionality. We thus present a novel sensing sleeve that uses textile electrodes to measure sEMG activity of intrinsic thumb muscles. We evaluate the sleeve's performance on detecting thumb movements and muscle activity during both isolated and isometric muscle contractions of the thumb and fingers. This work highlights the potential of textile-based sensors as a low-cost, lightweight, and non-obtrusive alternative to conventional sEMG sensors for wearable robotics. 
\end{abstract}


\section{Introduction}
Stroke-induced hemiparesis can lead to chronic upper limb weakness and reduced hand dexterity, which motivates developing effective tools to assist and rehabilitate these motor impairments \cite{barry}\cite{kamper}. Wearable robotic devices have emerged as solutions to assist stroke survivors in activities of daily life and provide tailored rehabilitation exercises for recovery. A key element of robotic assistance and therapy is continued, active use of the paretic hand \cite{hogan}. As stroke survivors engage their impaired muscles, surface electromyography (sEMG) signals from residual muscle activity can be used a control signal for a robotic orthosis \cite{meeker}.
\begin{figure}[t]
    \centering
    \includegraphics[width=\columnwidth]{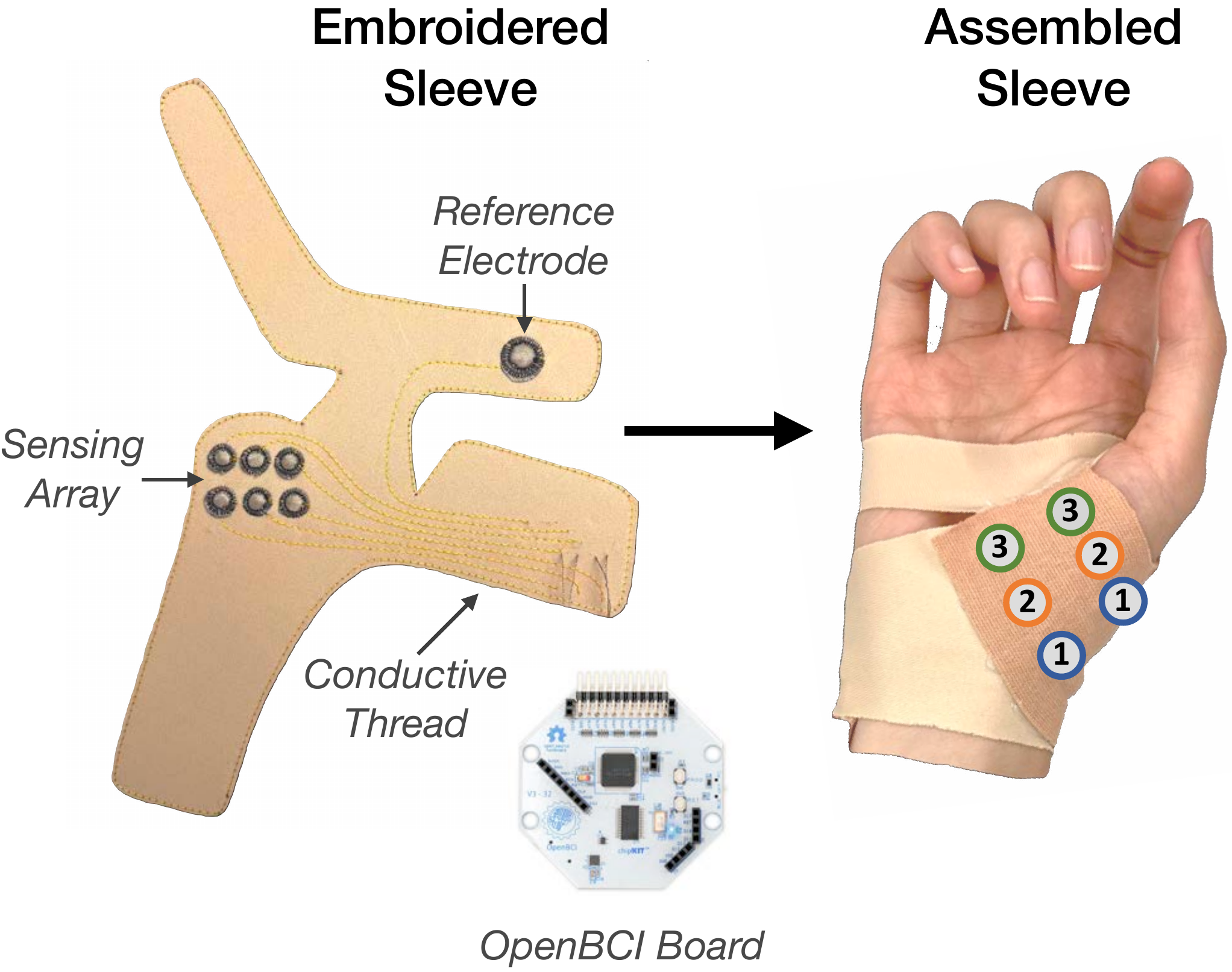}
    \caption{The fabric sensing sleeve. After embroidering the fabric electrodes and conductive thread, which act as wires to connect the sleeve to the OpenBCI board, the sleeve is assembled into a wearable form. The sensor array is positioned on the thenar eminence of the thumb to detect when the thumb moves into abduction.}
    \label{fig:fig_1}
\end{figure}

Such robotic devices commonly place sEMG sensors on the upper forearm to capture muscle activity from the hand extensor muscles (\textit{extensor digitorum}, ED) and hand flexor muscles \cite{zhiyuanlu}, and use machine learning models to infer user intent. Discerning user intent from forearm muscle signals, however, is challenging in stroke survivors due to abnormal muscle co-activation and neurological disorders \cite{miller}. Previous studies have employed multi-modal systems~\cite{tacca} that combine forearm sEMG signals with other sensing modalities to improve intent inferral in stroke survivors. 

The intrinsic muscles in the hand offer an alternate sensing region of the body for intent inferral. Muscles in the thenar eminence are responsible for some thumb movements during grasping and fine motor movements \cite{long}. Research has also suggested that in stroke survivors, the \textit{abductor pollicis brevis} (APB) in the thenar eminence has reduced spasticity compared to other muscles in the hand~\cite{towles}.  

However, despite these advantages, intrinsic hand muscles are underexplored as sensing regions for assistive robotics. In practice, sensing these muscles often requires bulky, adhesive sEMG electrodes specifically placed on the thenar eminence to capture activity~\cite{xlhu}\cite{miaochen}. Hu et~al. combined use of ED and APB as control signals for rehabilitation with in a post-stroke training device~\cite{xlhu}. For hand gesture recognition, Chen et~al. placed sEMG sensors on the extrinsic muscles and APB and achieved high classification accuracy~\cite{miaochen}. However, such sensors placed on the palm can interfere with grasping and other functional use of the hand.

Textile sensors, made from conductive fabrics and threads, have emerged as an alternative to conventional sEMG electrodes for a non-obtrusive, low-profile sensing interface. Finni et~al. embedded conductive fabric into sports shorts for measuring muscle activity during knee extension exercises~\cite{finni}. By sewing textile electrodes into fabric sleeves placed on the forearm, previous works have used these sensors to measure extrinsic muscle activity for hand gesture recognition. Kamavuako et~al. embroidered conductive thread with metal button connectors to measure forearm muscle activity for hand gesture classification~\cite{kamavuako}. Wang et~al. created an 8-channel fabric myoelectric armband to classify forearm actions~\cite{huiwang}. A textile sEMG sensor, however, has not been used for sensing intrinsic muscle activity in the hand.

In this paper, we introduce a textile sensing array embroidered in a fabric sleeve to measure sEMG activity of the intrinsic thumb muscles. The main contributions of the work are as follows:
\begin{figure}[t!]
    \centering
    \includegraphics[width=\columnwidth]{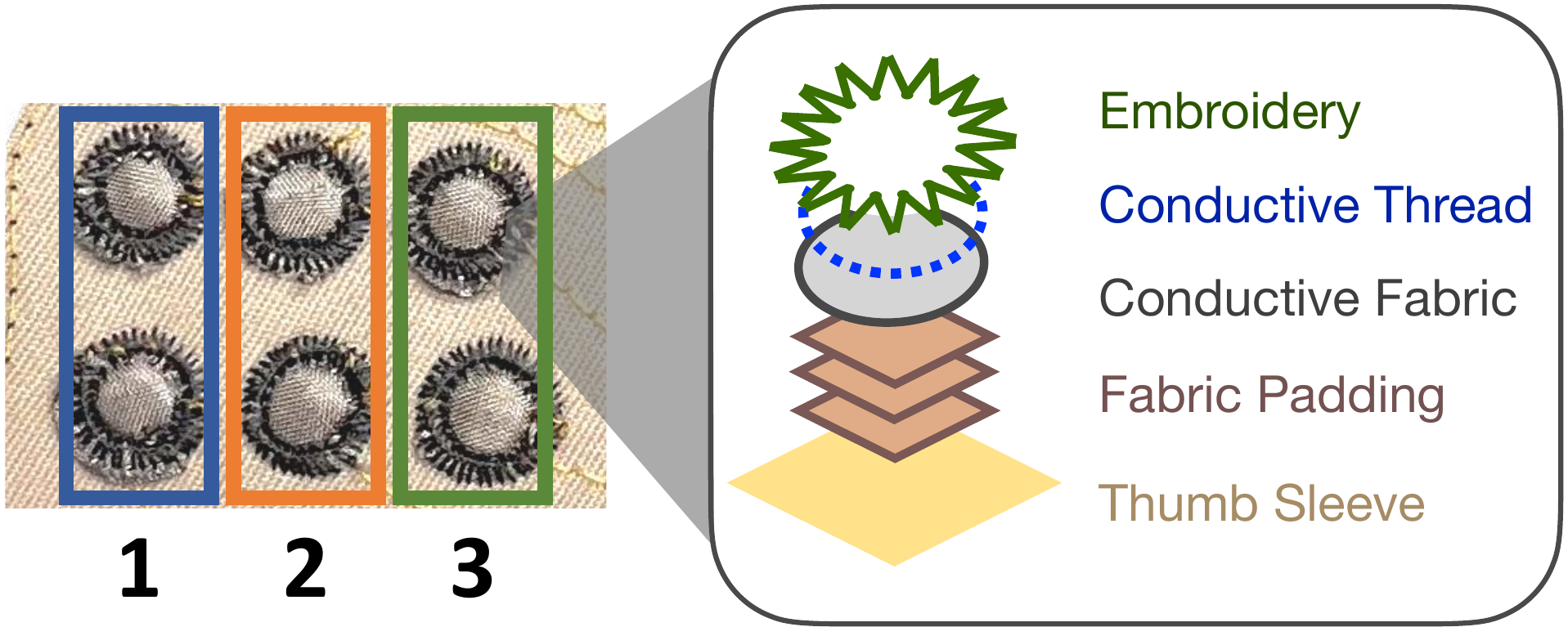} 
    \caption{The 3-channel textile sensing array, labeled by color and channel number. Each textile electrode is made of conductive fabric embroidered to a fabric substrate with internal padding to improve skin-electrode contact. Conductive thread is stitched around the perimeter of the electrode to connect the electrode to the bioamplifier board.}
    \label{fig:electrode}
\end{figure}

\begin{itemize}
    \item To the best of our knowledge, we are the first to use fabric sensors to measure surface EMG activity of intrinsic hand muscles. The fabric sensors are built with off-the-shelf fabrics and are optimized to mitigate motion artifacts and improve skin-contact quality. This sleeve does not require adhesive to apply and is adjustable to different hand sizes.
    \item We demonstrate that the textile sensors can distinguish thumb muscle activity from finger movements. In addition, training a classifier with only the 3-channel sensing array achieves promising classification accuracy for detecting hand opening and closing.
    \item Our results demonstrate the fabric sensing sleeve's potential as an effective sEMG sensor for intent inferral, highlighting the value of integrating both extrinsic and intrinsic muscle sensing modalities to improve model classification performance.
\end{itemize}

\section{Fabric Sensing Sleeve Design}

The fabric sleeve presented in this work is a prototype \mbox{3-channel,} fabric, sEMG sensing sleeve that measures the intrinsic muscle activity of the thumb. The materials for fabricating the sleeve are off-the-shelf and low-cost, which enables modular design for project-specific sensing array customization and scalability.

The biopotential sensing electrode is made of a layer of conductive fabric \cite{fabric} embroidered to an elastic fabric substrate. Three layers of fabric padding are placed between the conductive fabric and fabric substrate to increase electrode concavity and improve skin-electrode contact (Fig.~\ref{fig:electrode}). The reference electrode is 8 mm in diameter, and each sensing electrode is 6 mm in diameter, with a center-to-center electrode distance per channel of 9 mm. 

We designed the fabric sleeve to place the 3-channel sensing array over the thenar eminence of the palm, with the reference electrode placed on the back of the hand (Fig.~\ref{fig:fig_1}). Each fabric electrode routes to a bioamplifier board \mbox{(Cyton, OpenBCI)} \cite{openbci} via machine-sewn conductive thread \cite{thread}, which is positioned on the outside of the sleeve and insulated with sports kinesiology tape to reduce noise and prevent off-target skin contact. We confirmed the connectivity between the fabric electrode and the wiring adapter with a multimeter and measured an average resistance of $46.3\ \Omega \pm 3.3\ \Omega$ across all electrodes and their conductive thread wires.
\begin{figure}[t] 
    \centering
    \includegraphics[width=\linewidth]{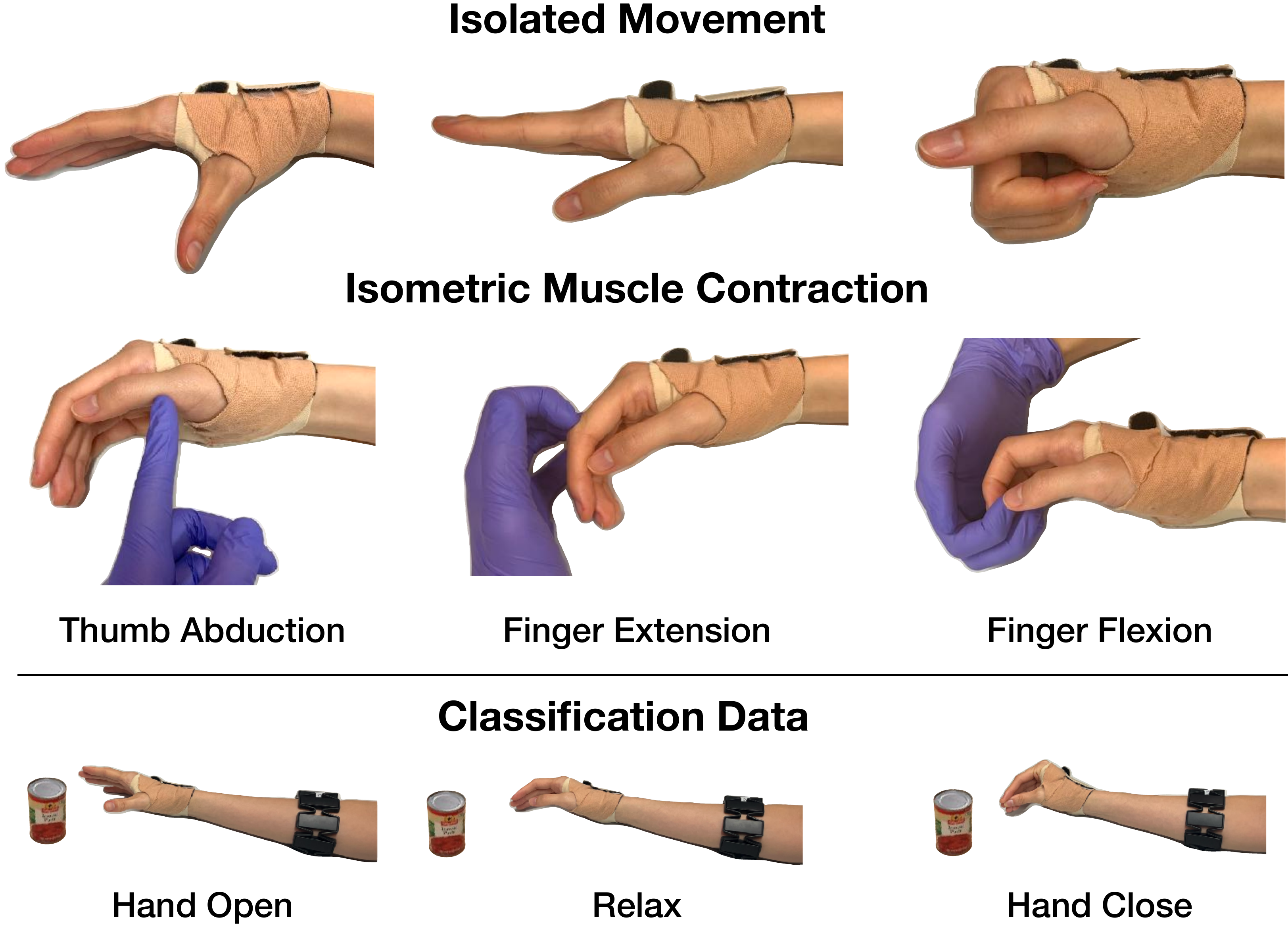} 
    \caption{Schematic of experimental protocol. During the isolated movement task, participants moved their thumb or fingers from rest to an engaged position as shown. The isometric muscle contraction task required participants to resist external force from the experimenter's hand to engage the relevant muscles for the movement. Classification data collection involved participants opening and closing their hands into a lumbrical grasp, as shown.}
    \label{fig:protocol}
\end{figure}

\section{Methods}

We designed two experiments to evaluate the fabric sleeve's performance on nine healthy participants and one stroke participant. Our first objective was to determine whether the sleeve could distinguish thumb-specific movements from overall finger movements and movement noise. The tasks chosen to evaluate this aspect included both dynamic and static thumb and finger movement conditions to isolate intrinsic thumb muscle activity. Our second objective was to explore whether signals measured by the fabric sleeve could be used for intent inferral. To test this, we trained a machine learning classifier to detect hand opening and hand closing from sEMG signals measured by the fabric sleeve.

\subsection{Isolated Finger Movement Task}
The Isolated Finger Movement Task distinguished between thumb-only movement and coordinated finger movement. Participants moved either their thumb or all four fingers from a resting position into an engaged position and held the position with moderate effort for five seconds (Fig.~\ref{fig:protocol}). To assess thumb-specific muscle activity, participants moved their thumb from a resting position into abduction. For finger-specific muscle activity, participants simultaneously moved all four fingers from a resting position into either extension or flexion.

\subsection{Isometric Muscle Contraction Task}
The Isometric Muscle Contraction Task targeted thumb-specific muscle activity in a static position. Participants performed the same actions as the isolated finger movement task but with added resistance from the experimenter (Fig. \ref{fig:protocol}). To minimize movement artifacts, participants simply resisted the external force for five seconds, without moving their thumb or fingers. For the thumb, the experimenter applied pressure with their finger to the lateral side of the participant's thumb MCP joint. For finger extension, the experimenter applied resistance on the dorsal side of their fingers at the head of the proximal phalanx. For finger flexion, the experimenter applied resistance on the palmar side of the participant's fingertips.

\subsection{Hand Gesture Classification}
To test whether signals from the fabric sleeve could be used for intent inferral, we collected data while participants performed a series of hand gestures. We instructed participants to open and close their hand as if they were to grasp a metal can, which we provided as a visual aid to encourage thumb abduction in a lumbrical grasp (Fig. \ref{fig:protocol}). We collected two datasets for training and testing, during which the participant opened and closed their hand 8 times, each for 5 seconds, with 5 seconds of rest between each movement.

\subsection{Data Collection and Analysis}
During data collection, participants were seated with their forearms resting on the table and elbows bent at a 90 degree angle. Participants donned both the fabric sensing sleeve and the commercial, 8-channel forearm sEMG armband \mbox{(Myo, Thalmic Labs)}. Before donning the fabric sensing sleeve, we prepped the participant's hand by wiping their skin with a moist towelette and adjusted the position of the sensing array to lie over the thenar eminence of the thumb.

Using the OpenBCI software, we measured the skin-electrode impedance to ensure sufficient contact and monitored the contact throughout the session. The raw voltage sEMG data from the fabric sleeve was passed through a 4th order Butterworth bandpass filter from 50 Hz to 115 Hz with a notch filter at 50 Hz and 60 Hz to eliminate environmental powerline noise. We collected sEMG data from the fabric sensing sleeve at 250 Hz and sEMG data from the Myo armband at 200 Hz.

\begin{figure}[t!] 
    \centering
    \includegraphics[width=0.95\linewidth]{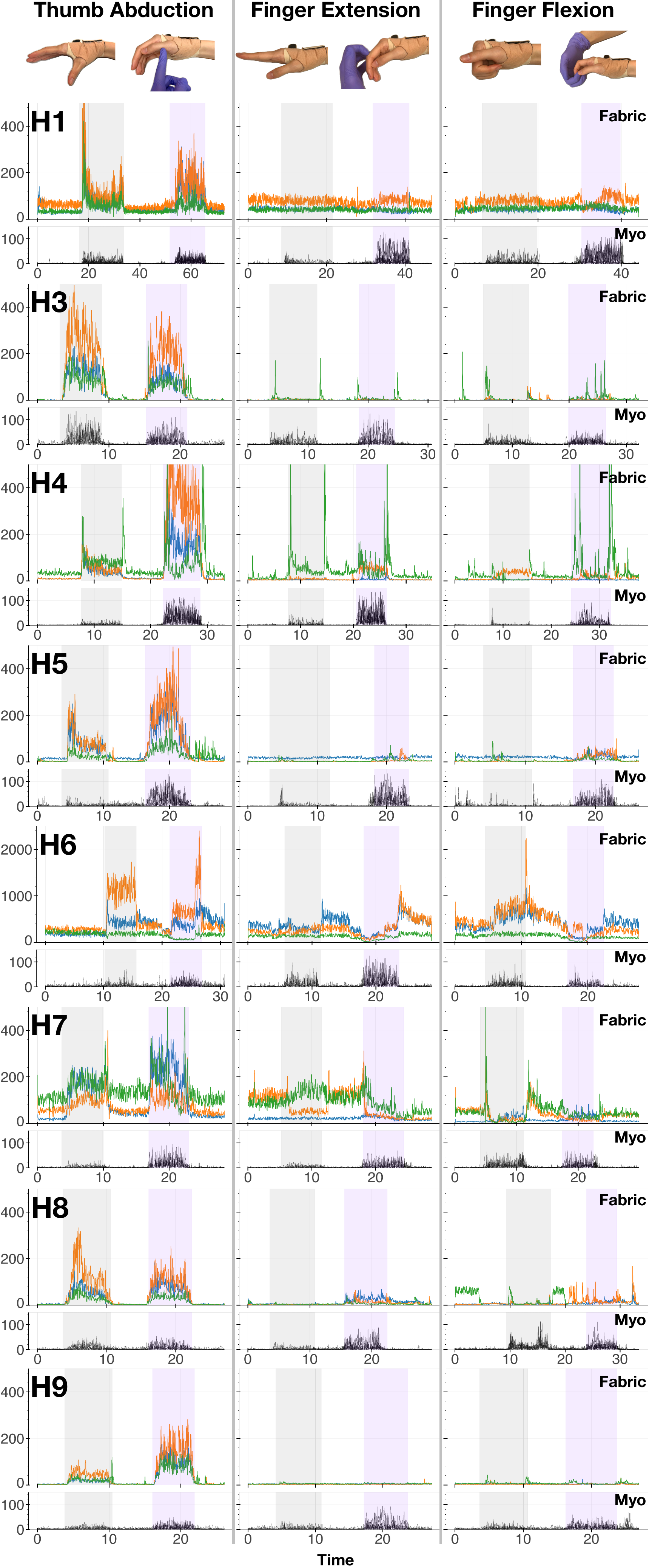} 
    \vspace{-3mm}
        \caption{(best seen in color) Data from our fabric sensing sleeve and commercial armband (Myo) during the isolated muscle movement and isometric muscle contraction tasks over time (s). Grey bars denote isolated movement, and purple bars denote isometric contraction. Each row depicts data from one healthy participant, and each column corresponds to a hand movement (Fig. \ref{fig:protocol}). For each figure, the top plot shows the fabric sensor activity in $\mu V_{rms}$ (blue = Channel 1, orange = Ch. 2, green = Ch. 3). Y-axis limit is $500\ \mu V$  except for H6 (2,500\ $\mu V$), and all cut-off peaks are below 1,000 $\mu V$. Bottom plot depicts corresponding Myo sensor activity in proprietary values (a.u.).}
    \label{fig:plots}
\end{figure}

For the Isolated Finger Movement and Isometric Muscle Contraction Tasks, we needed to to quantify changes in muscle activity between rest and active phases. To achieve this, we first rectified the raw EMG signal and passed it through a 40 Hz low-pass filter. Next, we found the average signal amplitude per sensing channel (3 channels in the fabric sensing sleeve and 8 channels in the commercial armband) to calculate the ratio of signal amplitude between the resting phase and the active phase of each of the two tasks.

For hand gesture classification, we processed the raw output from the OpenBCI board by extracting overlapping sliding windows of size 250 (1 second) with an offset of 10 samples. Each window was labeled with the ground truth at the end of the window, indicating the intent. From each window, we extracted both time-domain and frequency-domain features from each channel in each window, including Mean Absolute Value, Root Mean Square, Variance, Waveform Length, Zero Crossing Rate, Mean Frequency, Median Frequency, and Total Power. After feature extraction, we standardized features by removing the mean and scaling to unit variance using the training set.

We then trained three machine learning models to classify the extracted features: Linear Discriminant Analysis (LDA), Random Forest (RF), and a Multi-layer Perceptron (MLP). LDA was chosen for its simplicity and effectiveness with lower-dimensional data, RF for its ability to capture nonlinear relationships, and MLP for its capability to model more complex patterns. Each model was trained using subject-specific data, with distinct training and test sets.

\subsection{Participants}
We recruited nine healthy participants (five male, four female, ages 22-27)\footnote{H2 was later excluded from the study due to insufficient sEMG activity measured by the commercial armband.}. We also recruited a single stroke participant with chronic hemiparesis (male, right-hand impaired, scores of 1+ elbow flexors and 1 finger flexor on the Modified Ashworth Scale) for feasibility testing in a clinical setting under the supervision of an occupational therapist. Testing with all participants was approved by the Columbia University Institutional Review Board (IRB-AAAS8104).

\begin{table}[t!]
    \renewcommand{\arraystretch}{1.3} 
    \caption{Average Change in Signal Amplitude of Fabric Sensing Sleeve Relative to Relaxed State}
    \vspace{-2mm}
    \label{tab:ratio_table}
    \centering
    \resizebox{\linewidth}{!}{
    \begin{tabular}{c|ccc|ccc}
    \toprule
    \multirow{2}{*}{\textbf{Participant}} & \multicolumn{3}{c|}{\textbf{Isolated Finger Movement}} & \multicolumn{3}{c}{\textbf{Isometric Muscle Contraction}} \\
    \cline{2-7}
    & \textbf{Thumb} & \textbf{FE} & \textbf{FF} & \textbf{Thumb} & \textbf{FE} & \textbf{FF} \\
    \midrule
    H1 & \textbf{2.2} & 1.0 & 1.1 & \textbf{2.4} & 0.9 & 1.2 \\
    H3 &  \textbf{88.5} & 4.1 & 2.1 & \textbf{66.6} & 2.9 & 2.6 \\
    H4 &  \textbf{4.9} & 2.5 & 2.4 & \textbf{18.9} & 5.7 & 2.5 \\
    H5 &  \textbf{13.2} & 1.2 & 1.1 & \textbf{30.9} & 3.5 & 4.8 \\
    H6 &  \textbf{2.4} & 1.0 & 1.5 & \textbf{1.9} & 0.7 & 0.5  \\
    H7 &  \textbf{3.7} & 1.0 & 1.4 & \textbf{5.2} & 0.8 & 1.6 \\
    H8 & \textbf{ 20.5} & 0.8 & 0.9 & \textbf{22.0} & 4.3 & 2.1 \\
    H9 & \textbf{11.6}& 1.2 & 1.4 & \textbf{39.3} & 1.2 & 1.2 \\
    \cline{1-7}
    Avg & \textbf{18.4} & 1.6 & 1.5 & \textbf{23.4} & 2.5 & 2.1\\
    \bottomrule
    \end{tabular}}
    \vspace{-2mm}
\end{table}
\section{Results \& Discussion}

\subsection{Isolated Muscle Movement Results}

We evaluated the sensor performance by finding the ratio between the average signal amplitude during the resting phase and the active phases. The raw data is visualized in Fig.~\ref{fig:plots}, and Table \ref{tab:ratio_table} shows the average ratio values across all sensing channels measured by the fabric sleeve. In Table \ref{tab:ratio_table}, the hand movement labels are abbreviated as follows: Thumb for thumb abduction, FE for finger extension, and FF for finger flexion.

During the Isolated Muscle Movement Task, we observed the largest increase in signal amplitude during thumb abduction. For the thumb abduction condition, we report an average increase in signal amplitude by a factor of 18.4 from the relaxed position across all healthy participants. In contrast, the average signal change for the finger conditions were much smaller, with a factor of 1.6 for finger extension and 1.5 for finger flexion. 

The difference in average signal magnitude between the thumb condition and the finger conditions demonstrated that the fabric sensing sleeve measured high activity during thumb activity and low activity when thumb movement was absent. Even during finger movement, the fabric sensors detected low activity as long as the thumb was not engaged. 

To confirm the low activity detected by the fabric sensors during finger movement was not due to a lack of effort, we recorded the commercial sEMG armband values of concurrent extrinsic muscle activity in the forearm during finger extension and flexion (Fig. \ref{fig:plots}). The sustained peaks in signal activity measured by the commercial sEMG armband suggested greater activation of the extrinsic muscles than the intrinsic muscles. During the isolated finger extension and flexion tasks, the fabric sensor did not detect sustained activation during the 5 second period, indicating that the sensor was more sensitive to thenar eminence muscle activity than to finger movements. 

We observed that three healthy participants had transient peaks at the start and end of each movement. However, the lack of sustained activation as observed in the thumb movement tasks supported the conclusion that the sensor primarily detected thumb muscle activity. These transient peaks could also be attributed to the participant's ability to maintain a relaxed thumb position during both the isolated finger movements.

\subsection{Isometric Muscle Contraction Results}

The Isometric Muscle Contraction Task focused on measuring changes in thenar eminence muscle activity without finger movement to mitigate potential motion noise. With isometric muscle contraction alone, the results from this task aligned with those of the isolated muscle movement task–the thumb abduction condition exhibited the largest change in average signal amplitude from the relaxed phase. Across the healthy participants, we report an average increase in signal amplitude by a factor of 23.4 for the thumb abduction condition, compared to a factor of 2.5 and 2.1 increase from relaxed during the finger extension and flexion conditions, respectively. 

During the thumb abduction condition, a visible and sustained peak in the signal during both isolated movement and isometric muscle contraction indicated its sensitivity to thumb activity, whether from movement or muscle engagement \mbox{(Fig \ref{fig:plots})}. While muscle shape changes during contraction could affect the contact surface with the sensors, the sustained peak during isometric muscle contraction—where no thumb movement occurred–indicated that the signal was likely due to muscle activation rather than motion artifacts.

A notable outlier in signal quality from Fig. \ref{fig:plots} was H6, whose baseline signal and overall signal amplitude was larger than all other participants. As a simple proxy for thenar eminence size, we measured the depth of the thenar eminence for each healthy participant between the palmar and dorsal sides of the hand. H6 had a thenar eminence depth of 25 mm, compared to the average thenar eminence depth of $34 \pm 3.9 $ mm among all other healthy participants. Furthermore, we observed that H6 had drier skin compared to other participants. Thus, H6's hand size and skin condition could have affected skin-electrode contact with the fabric sleeve for reduced performance.

\subsection{Classification Performance}
Having collected simultaneous fabric sensor and commercial sEMG armband data during classifier data collection, we trained three machine learning models (LDA, RF, and MLP) with the fabric sensor and the commercial armband separately to assess their classification accuracy. The results are shown in Table \ref{tab:classification_table}, with the highest classification accuracy for the fabric sensor and commercial armband for each participant marked in bold. For the fabric sensor, the highest classification accuracy of 92.6\% was achieved with the LDA model. The commercial sEMG armband achieved a maximum classification accuracy of 91.1\% with the RF model. 

The classification accuracy of the fabric sensing sleeve highlighted its potential as a viable sensor for intent inferral. For half of the participants, the 3-channel fabric sensing sleeve achieved a maximum classification accuracy within 5\% of the commercial armband. This comparison is not intended to suggest the fabric sensing sleeve could replace the commercial forearm sEMG sensor. Rather, this suggests that sEMG signals from thenar eminence activity alone captured sufficient signal information for intent inferral. These results suggest the fabric sensing sleeve could augment a forearm sEMG sensing setup to improve intent inferral.

\begin{table}[t]
    \renewcommand{\arraystretch}{1.3} 
    \caption{Classification Accuracy of Fabric Sensing Sleeve and Myo Armband across 3 candidate models}
    \vspace{-2mm}
    \label{tab:classification_table}
    \centering
    \resizebox{\linewidth}{!}{ 
    \begin{tabular}{c|ccc|ccc}
    \toprule
    \multirow{2}{*}{\textbf{Participant}} & \multicolumn{3}{c|}{\textbf{Fabric Sensor}} & \multicolumn{3}{c}{\textbf{Myo Armband}}\\
    \cline{2-7}
    & \textbf{LDA} & \textbf{RF} & \textbf{MLP} & \textbf{LDA} & \textbf{RF} & \textbf{MLP} \\
    \midrule
    H1 & \textbf{70.1\%} & 68.6\% & 65.1\% & 84.6\% & \textbf{84.7\%} & 80.4\% \\
    H3 & \textbf{84.7\%} & 83.8\% & 79.4\% & \textbf{84.1\%} & 83.1\% & 82.1\% \\
    H4 & \textbf{84.4\%} & 80.8\% & 83.0\% & \textbf{86.6\%} & 86.3\% & 81.5\% \\
    H5 & \textbf{81.9\%} & 69.6\% & 66.7\% & 78.7\% & \textbf{84.8\%} & 79.0\% \\
    H6 & 13.0\% & \textbf{67.5\%} & 42.0\% & 77.3\% & \textbf{81.3\%} & 74.6\% \\
    H7 & 70.1\% & 46.5\% & \textbf{70.6\%} & 88.2\% & \textbf{91.1\%} & 89.9\% \\
    H8 & \textbf{92.6\%} & 91.1\% & 87.7\% & 89.3\% & \textbf{90.7\%} & 90.0\% \\
    H9 & 77.5\% & 78.6\% & \textbf{80.3\%} & 86.8\% & \textbf{87.4\%} & 83.3\% \\
    S1 & 68.6\% & \textbf{74.5\%} & 68.7\% & 83.2\% & 83.1\% & \textbf{86.7\%} \\
    \bottomrule
    \end{tabular}}
    \vspace{-2mm}
\end{table}

\subsection{Preliminary Testing with Stroke Participant}

We performed a modified version of this protocol with one stroke survivor who had limited active thumb movement. We collected isolated thumb movement and isometric thumb muscle contraction data and observed a 0.8 fold change in signal amplitude from resting during the isolated thumb movement and 0.8 fold change during the isometric thumb muscle contraction. Unlike the healthy participants, we did not observe increased signal amplitude with the fabric sleeve during thumb engagement. However, the participant had a 3.9 fold increase in signal measured by the commercial sEMG armband during isolated muscle movement, and a 2.6 fold increase in signal during isometric muscle contraction. The lack of signal activity measured by the fabric sleeve could be because the stroke survivor had difficulty isolating the fine-motor movement of the thumb from overall hand engagement. 

When asked to engage their whole hand, the fabric sleeve captured some signals from the stroke participant. We collected two datasets, during which the participant was instructed to open and close their hand 3 times, each for 5 seconds, with 5 seconds of rest between each movement. The fabric sensor had a maximum classification accuracy of 74.5\% with the RF model, and the commercial sEMG armband had a maximum classification accuracy of 86.7\% with the MLP model. While the fabric sleeve may not distinguish isolated thumb movement in stroke survivors, this result shows potential for its use with a commercial armband for intent inferral. 

\subsection{Future Work \& Limitations}
The modularity of the textile sensor design presents an opportunity to integrate additional, targeted sEMG sensors on the body for improved intent inferral. A future direction of this work would be integrating additional textile sensors to create an entirely fabric-based sensing sleeve to measure intrinsic and extrinsic muscle activity. Developing a multi-modal machine learning model to leverage both intrinsic and extrinsic muscles would be a logical next step toward improving classification accuracy for intent inferral.  

Limitations of this work include the small sample size of stroke participants. For the stroke participant, we could not rule out that the activation may have been from \textit{flexor pollicis brevis} rather than \textit{abductor pollicis brevis}, as stroke survivors may have difficulty distinguishing between these motions. Expanding the number of stroke participants in the work would allow us to draw more conclusions on using intrinsic hand muscle activity for intent inferral and the use of textile sensors to detect abnormal muscle signals from impaired individuals. 

Further research will focus on characterizing the textile sensors, including a comparison of its performance to commercial sEMG systems on the thenar eminence and evaluating the sensors' dynamic response and skin-electrode contact quality. As seen with participant H6, two factors--–hand size and skin condition--–motivates creating different fabric sleeve sizes and testing sEMG preparation gel to improve skin-electrode contact. Exploring the impact of different skin conditions and sEMG preparation gel could help address the fluctuations in signal quality observed among healthy participants to ensure better sensor performance.

\section{Conclusion}
 In this work, we present a fabric sleeve with embedded textile sEMG sensing array to measure thumb movement and muscle activity of intrinsic hand muscles. To the best of our knowledge, this is the first fabric sensing sleeve that uses textile sensors to measure intrinsic hand muscle sEMG activity. With a three-channel sensing array on the thenar eminence, we demonstrate that the textile sensors can effectively discriminate intrinsic thumb movements from extrinsic finger movements during both Isolated Finger Movement and Isometric Muscle Contraction Tasks. Furthermore, we show that the fabric sleeve alone can achieve promising classification accuracy in classifying hand open and hand close poses for intent inferral. In the context of assistive robotics, where wearability and comfort are key aspects of device design, textile sensors offer a non-obtrusive interface for muscle sensing. Textile sensors for sEMG sensing is a low-cost, lightweight, and scalable technology that motivates future work in its improvement and adoption in wearable devices.


\bibliography{newpaperpile.bib} 
\end{document}